\documentclass[prx,twocolumn,superscriptaddress,tightenlines,longbibliography]{revtex4-1}

\usepackage{color}
\usepackage{amssymb}
\usepackage{graphicx}
\usepackage{xcolor}
\usepackage[normalem]{ulem}

\definecolor{darkgreen}{RGB}{89, 175, 99}

\definecolor{grey}{rgb}{.6,.6,.6}

\begin{document}

\title{What limits the simulation of quantum computers?}

\author{Yiqing Zhou}
\affiliation{Department of Physics, University of Illinois at Urbana-Champaign, Urbana, IL 61801, USA}
\affiliation{Center for Computational Quantum Physics, Flatiron Institute, New York, NY 10010, USA}

\author{E. Miles Stoudenmire}
\affiliation{Center for Computational Quantum Physics, Flatiron Institute, New York, NY 10010, USA}

\author{Xavier Waintal}
\affiliation{Univ. Grenoble Alpes, CEA, IRIG-Pheliqs, 38054 Grenoble, France}
\date{\today}

\begin{abstract}
An ultimate goal of quantum computing is to perform calculations beyond the reach of any classical computer. It is therefore imperative that useful quantum computers be very difficult to simulate classically; otherwise classical computers could be used for the applications envisioned for the quantum ones. {\it Perfect} quantum computers are unarguably exponentially difficult to simulate: the classical resources required grow exponentially with the number of qubits $N$ or the depth $D$ of the circuit. 
This difficulty has triggered recent experiments on deep, random circuits that aim to demonstrate that quantum devices may already perform tasks beyond the reach of classical computing. These {\it real} quantum computing devices, however, suffer from many sources of decoherence and imprecision which limit the degree of entanglement that can actually be reached to a fraction of its theoretical maximum. 
They are characterized by an exponentially decaying fidelity $\mathcal{F} \sim (1-\epsilon)^{ND}$ with an error rate $\epsilon$ per operation as small as $\approx 1\%$ for current devices with several dozen qubits or even smaller for smaller devices.

In this work, we provide new insights on the computing capabilities of real quantum computers by demonstrating
that they can be simulated at a tiny fraction of the cost that would be needed for a perfect quantum computer. 
Our algorithms compress the representations of quantum wavefunctions using matrix product states (MPS),
which are able to capture states with low to moderate entanglement very accurately.
This compression introduces a finite error rate $\epsilon$
so that the algorithms closely mimic the behavior of real quantum computing devices. 
The computing time of our algorithm
increases only linearly with $N$ and $D$ in sharp contrast with exact simulation algorithms.
We illustrate our algorithms with simulations of random circuits for qubits connected in both one and two dimensional lattices.
We find that $\epsilon$ can be decreased at a polynomial cost in computing power down to a minimum error $\epsilon_\infty$. Getting below $\epsilon_\infty$ requires computing resources that increase exponentially with $\epsilon_\infty/\epsilon$.
For a two dimensional array of $N=54$ qubits and a circuit with Control-Z gates, error rates better than state-of-the-art devices can be obtained on a laptop in a few hours. For more complex gates such as a swap gate followed by a controlled rotation, the error rate increases by a factor three for similar
computing time.
Our results suggest that, despite the high fidelity reached by quantum devices, only a tiny fraction $(\sim 10^{-8})$ of the system Hilbert space is actually being exploited.
\end{abstract}

\maketitle

\section{Introduction}
Operating a quantum computer is a race against the clock. The same phenomenon enabling the potential computing power of quantum computers---entanglement---is also responsible for decoherence when it occurs with unmonitored degrees of freedom. The main challenge of quantum computing is to quickly build entanglement between the qubits before imperfections or decoherence overly corrupt the quantum state. 
This decoherence is an intrinsic characteristic of any quantum computer and its origin and consequences must be understood thoughtfully. But in all hardware realizations, it means each operation incurs a loss of fidelity relative to the ideal target quantum state.

As different experimental platforms for quantum manipulation make rapid, impressive advances, there has been a justifiable interest in the computational capability of near-term quantum computers \cite{Preskill2018}. One of the key questions is when and how to achieve the goal of ``quantum supremacy'' \cite{Preskill2012},
which is the crossover point where a quantum system ceases to be within reach of simulation on a classical computer. Precise circuits and fidelity metrics
have been designed to meet this goal \cite{Boixo2018}. 
Recently, an experiment using $N=53$ qubits and a circuit of depth $D=20$ has reached a multi-qubit 
fidelity ${\cal{F}} = 0.002$ \cite{Arute2019}. According to the authors, such an experiment would take thousands of years to be simulated on the largest existing supercomputers. This statement was then challenged by another estimate which claims that only two days would be needed \cite{Pednault2019}. Such a disparity between estimates raises the question of the difficulty of simulating a quantum computer and consequently of the true computing power realized in a quantum computer.

The implicit assumption behind quantum supremacy as well as the most appealing applications of quantum computing is that a quantum computer is exponentially hard to simulate. Indeed, in recent years many techniques have been developed to simulate quantum computers, and they all have an exponential cost in some parameter. A brute force approach where one holds the full quantum state in memory as a large vector of size $2^N$ ($N$: number of qubits) requires a computing time and memory that scales exponentially with $N$ but linearly with the depth $D$ of the circuit. Other approaches require a computing time that scale exponentially with the number of two-qubit gates \cite{Vidal2003,Chen2018,Guo2019,Pan2019}, with the number of non-Clifford gates \cite{Aaronson2004} and/or with the number of gates that are non-diagonal in a chosen basis \cite{Boixo2017,Jnsson2018}. 
All these techniques can simulate {\it perfect} quantum computers.
In all cases, the required computing resources are exponential so that getting beyond $N=50$ and a depth $D=20$ for an arbitrary circuit is extremely difficult. 

In this article, we show that {\it real} quantum computers can be simulated at a tiny fraction of the cost that would be needed for a {\it perfect} quantum computer. 
To do so, we take advantage of the fact that in real quantum computers, decoherence limits the amount of entanglement that can be built into the quantum state to a fraction of what the exponentially large Hilbert space would suggest.
Our algorithms use a compressed wavefunction representation that achieves very high accuracy
for states with low to moderate entanglement.
This compression introduces a finite error rate $\epsilon$ per two-qubit gate.
Hence, in this class of algorithms the limiting factor is the fidelity with which
the calculation is performed while the computing time is linear in both the number of qubits $N$ and the depth $D$. These algorithms ``mimic'' actual quantum computers both in the sense of how they scale with $N$ and $D$, and in the sense that the main difficulty lies in increasing the fidelity of the calculation: a small finite error $\epsilon$ is made each time a two-qubit gate is applied to the state. Therefore, they offer a better reference point than exact simulation algorithms for assessing the computing power harvested by actual quantum chips. 

Our algorithms are based on tensor networks and more precisely on matrix product states (MPS) \cite{Schollwock2011}. 
MPS have been recognized very early as an interesting parameterization of 
 many-qubit quantum states for quantum simulations \cite{Vidal2003} and its generalizations are used in some of the most advanced quantum simulation approaches \cite{Markov2008}. However, so far, the focus of classical simulations of quantum hardware has been building essentially exact simulations techniques and little attention has been devoted to approximate techniques. Interestingly these exact techniques can require one to go well beyond double precision calculations \cite{Saitoh2013} which already hints at the link between error rate and underlying computing difficulty. 
 
The historical success of MPS has {\it not} been for exact calculations but, in contrast, for the development of controlled, approximate techniques to address quantum many-body physics problems. This includes the celebrated density matrix renormalization group (DMRG) algorithm \cite{White1992} which has provided precise solutions to a number of one-dimensional and quasi-one-dimensional problems, as well as time-dependent extensions \cite{Paeckel2019} and generalizations to higher dimensions through projected entangled pair states (PEPS) \cite{Verstraete2008} or multi-scale entanglement renormalization ansatz (MERA) \cite{Vidal2007} tensor networks. At the root of these successes is the fact that MPS naturally organizes states according to the amount of entanglement entropy between different parts of the system. Hence, slightly entangled systems can be easily represented with MPS. As entanglement entropy grows, one eventually truncates the basis. The associated error can be made arbitrarily small by keeping a larger set of basis states. 

In this article, we construct such an approximate technique in the context of quantum computing. Our chief result is that, for fidelities comparable to those reached experimentally, the computational requirement for simulating an imperfect quantum computer is only a tiny fraction of the requirements for a perfect one.

\section{Possible strategies for approximate simulations of quantum circuits}

Let us start by discussing possible strategies for simulating quantum circuits in an approximate manner.
Suppose that we have partitioned the qubits into two different sets $A$ and $B$ with 
respectively $N_A$ and $N_B$ qubits ($N_A + N_B = N$). Let us consider the two-qubit
gates that connect A and B and ignore gates internal to A or B. Performing a singular value decomposition (SVD) of such a gate, it can be written as a sum of terms that act separately on A and B. This sum contains two terms for the case of  usual gates (Control-NOT and Control-Z) and at most four terms for an arbitrary two-qubit gate. It follows that computing the state after $n$ of these gates amounts to keeping track of $2^n$ (up to $4^n$) different amplitudes. These amplitudes are the discrete analogue of 
Feynman paths and are referred to as such in the literature. For the random circuits that will be considered in this article, these $2^n$ amplitudes have essentially random phases. It follows that if one keep track of just a \emph{single} path, one reaches an overall multi-qubit fidelity ${\cal F} = (1/2)^n$ (or ${\cal F}=(1/4)^n$ in the worst situation). 
This very simple strategy could be used to simulate an arbitrary large number of qubits with low fidelity
per gate in a computing time $\sim n$. However, if one wants to keep a fixed fidelity per gate $f$ defined as ${\cal F} = f^n$, in analogy with real quantum computers, the number of paths $N_{\rm path}$ that must be tracked during the simulation is $N_{\rm path} = (2f)^n$, and hence increases exponentially with $n$. Such a strategy has been used in 
Ref.~\onlinecite{Arute2019} to validate the experimental results reported there.

We now seek algorithms where a constant fidelity $f$ can be obtained at a constant computing cost per gate, independent of the total number of gates $n$. One starts by writing a general state for the bipartite system as
\begin{equation}
|\Psi\rangle = \sum_{a,b} \Psi_{ab}|a\rangle_A|b\rangle_B
\end{equation}
where the states $|a\rangle$ ($|b\rangle$) form an orthonormal basis of A (B). Performing
a singular value decomposition (SVD)
\begin{equation}
 \Psi_{ab} = \sum_\mu U_{a\mu} S_\mu V_{\mu b},
\end{equation}
one can define an orthonormal basis
\begin{equation}
|\mu\rangle_A = \sum_{a} U_{\mu a}|a \rangle_A
\end{equation}
(with similar notation for the B subsystem) and arrive at the usual Schmidt decomposition of 
$|\Psi\rangle$:
\begin{equation}
|\Psi\rangle = \sum_{\mu} S_\mu|\mu\rangle_A|\mu\rangle_B
\end{equation}
in terms of a finite number of singular values $S_\mu$. States with only one non-zero singular value $S_0=1$
are simple, unentangled product states. A measure of the number of significant singular values needed to describe the state to high accuracy is given
by the entanglement entropy 
\begin{equation}
S = - {\rm Tr} \ \rho_A \log \rho_A = - {\rm Tr} \ \rho_B \log \rho_B = -\sum_\mu S^2_\mu\log S^2_\mu
\end{equation}
where $\rho_A$ ($\rho_B$) is the reduced density matrix for the subsystem A (B). The general strategy of DMRG-like algorithms is to keep only a finite number $\chi$ of the singular values. After a two-qubit gate that connects A and B, one performs a SVD decomposition of $\Psi_{ab}$ and truncate the state by keeping only the $\chi$ largest singular values. When $\chi \gg e^S $ this procedure is essentially exact. As the entanglement increases, this procedure lead to a  certain fidelity per gate $f<1$ that can be controlled by increasing the parameter $\chi$. Of interest to the present article is the typical value of $f$ that can be reached in a reasonable computing time.                                                                                                     

\section{Noisy algorithm in one dimension}

Above we motivated the truncated SVD of a two-qubit wavefunction as an approximation strategy
that works well for wavefunctions with only a moderate amount of entanglement. A natural generalization
of this strategy to the $N$-qubit case is to use matrix product states (MPS), which can be viewed
as a simultaneous Schmidt decomposition of the wavefunction across $N$ different partitions \cite{Vidal2003}
or equivalently a sequence of compatible SVD factorizations of the wavefunction, grouping
qubits $1,2,\ldots,j$ and $j+1,\ldots,N$ and performing an approximate SVD of the resulting 
matrix \cite{Schollwock2011}.

\label{sec:1D}
\subsection{MPS representation of the state}
We first consider a one dimensional network of $N$ qubits where two-qubit gates can be only applied directly between nearest neighbors. (Within this connectivity, gates acting on other non-neighboring qubits are still possible at the cost of using $\sim N$ SWAP operations to bring the qubits onto neighboring sites.) We define our MPS state in terms of $N$ tensors $M(n)$ as
\begin{widetext}
\begin{equation}
\label{eq:mps}
|\Psi\rangle = \sum_x \Psi_x |x\rangle = \sum_{i_1...i_N}\sum_{\mu_1...\mu_{N-1}} M(1)^{i_1}_{\mu_1} M(2)^{i_2}_{\mu_1\mu_2}
M(3)^{i_3}_{\mu_2\mu_3} \ldots  M(N)^{i_{N}}_{\mu_{N-1}}|i_1 i_2 i_3 \ldots i_N \rangle
\end{equation}
\end{widetext}

\begin{figure}[b]
    \centering
    \includegraphics[width=0.85\columnwidth]{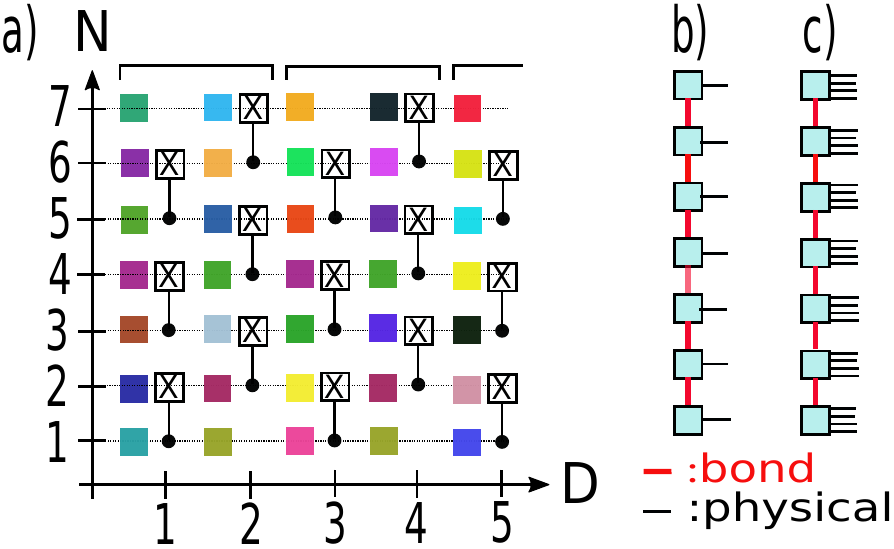}
    \caption{a) Sketch of the quantum circuit with $N$ qubits. The colored squares indicate arbitrary one-qubit gates while the dots connected to a cross indicate a two qubit gate such as Control-NOT or Control-Z. The depth $D$ counts the number of two-qubit gates performed in the sequence. b) structure of the Matrix Product States (MPS) for 1D circuits. Red lines indicate bond (or virtual) indices while thin black lines correspond to  physical indices. c) MPS structure for quasi-one dimensional structures.}
    \label{fig:Sketch}
\end{figure}
\noindent where the ``physical'' indices $i_n \in \{0,1\}$ span the $2^N$ dimensional Hilbert space while the
bond (or virtual) indices $\mu_n\in \{1,...,\chi_n\}$ control the maximum degree of entanglement allowed by the MPS.
$|x\rangle$ is a shorthand for $|i_1i_2...i_N\rangle$. 
If the $\chi_n$ are allowed to grow exponentially large as a function of $N$, then the MPS form of the wavefunction becomes exact and can represent any wavefunction \cite{Schollwock2011}.
In contrast, we will enforce $\chi_n\le\chi$ in what follows so that the resulting MPS represents an
approximation of the true wavefunction. The parameter $\chi$ controls the error rate made by our algorithm
as well as the computational and memory costs required to run it. As we will see below, applying a 
two-qubit gate takes $\sim \chi^3$ operations and the overall memory footprint is $N\chi^2$.
A sketch of the MPS structure is shown in Fig.~\ref{fig:Sketch}b. 

\begin{figure}[t]
    \centering
    \includegraphics[width=0.9\columnwidth]{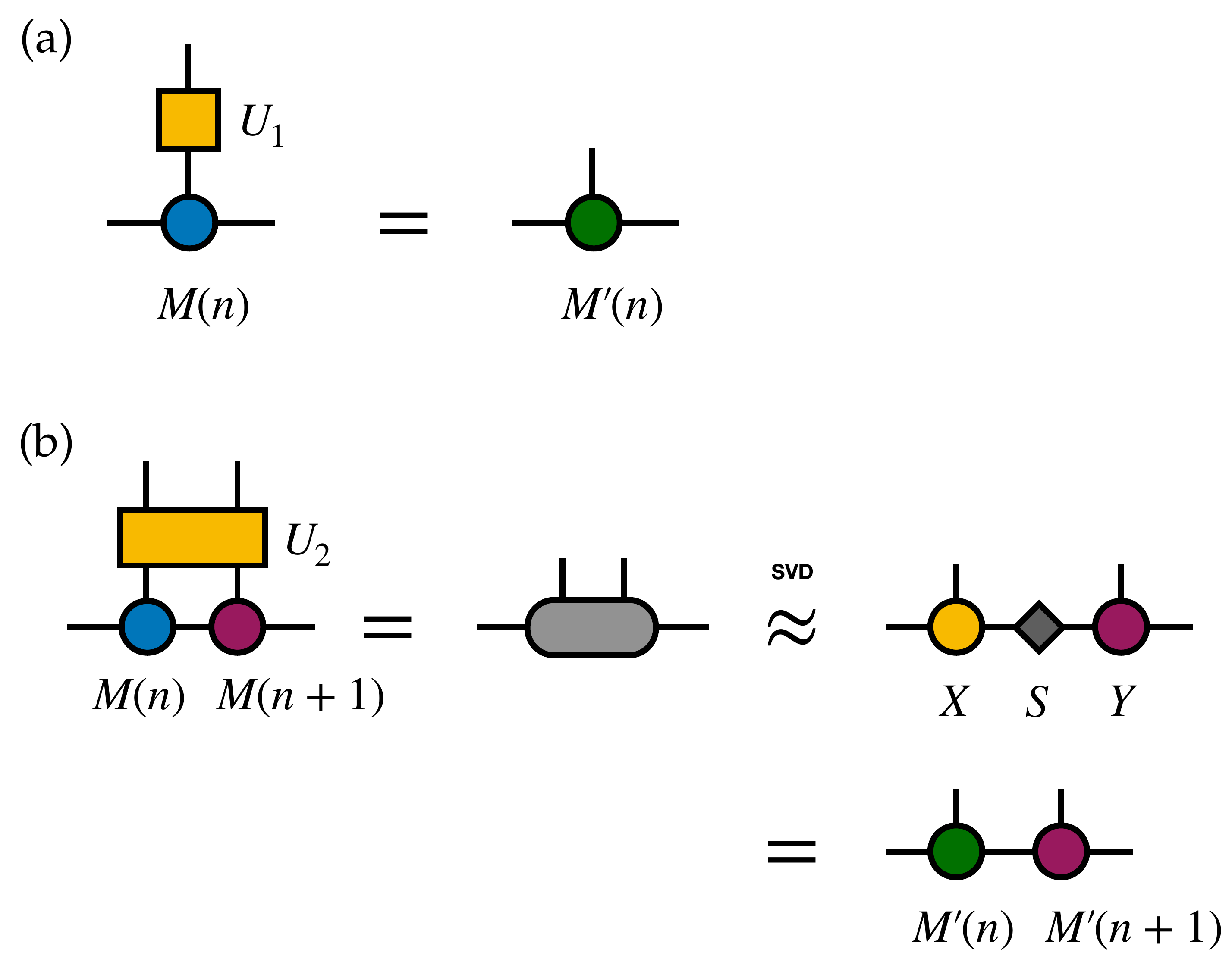}
    \caption{(a) Applying a single qubit gate  to an MPS can be done without approximation by multiplying the gate by a single MPS tensor. (b) To apply a two-qubit gate to qubits $n$ and $n+1$, one contracts the corresponding tensors together, then applies the gate. To restore the MPS form, the resulting tensor is decomposed with an SVD truncated to keep the largest $\chi$ singular values, and the matrix of singular values is multiplied into one of the unitary factors $X$ or $Y$. }
    \label{fig:single_qubit_mps}
\end{figure}

To be acceptable, our algorithm must provide the same features that a real quantum computer would provide. Applying a one-qubit gate $U$ on qubit $n$ can be done exactly and without increasing any of the $\chi_n$: it simply amounts to updating the corresponding tensor $M(n)\rightarrow M'(n)$:
\begin{equation}
M'(n)^{i'_n}_{\mu_{n-1}\mu_{n}}(n)  = \sum_{i_n} U_{i'_ni_n} M(n)^{i_n}_{\mu_{n-1}\mu_{n}}.
\end{equation}
as shown in Fig.~\ref{fig:single_qubit_mps}(a).
Calculating the overlap between different MPS states or calculating individual wavefunction amplitudes
$\langle i_1i_2...i_{N-1}i_N |\Psi\rangle$ can be done with contraction algorithms which, for MPS, can be done exactly in $\sim N\chi^3$ operations (see e.g. \cite{Schollwock2011} for a detailed description of standard MPS algorithms). It follows that one can also sample from the distribution $|\langle i_1i_2...i_{N-1}i_N |\Psi\rangle|^2$ within the same complexity. Quantum measurements (sampling of a given qubit followed by its projection) can also be done efficiently in a straightforward manner \cite{Ferris:2012}. 

To perform a two-qubit gate $U$ between qubit $n$ and qubit $n+1$, one first 
transforms the MPS into the so-called ``canonical form'' centered around the qubits of interest, 
through a series of $QR$ factorizations \cite{Schollwock2011}. This step is crucial for the accuracy of 
truncations of the MPS. 
The steps to apply the gate are then shown in Fig.~\ref{fig:single_qubit_mps}(b).
One first forms the two-qubit tensor 
\begin{equation}
\label{eq:1D:1}
T^{i_{n}i_{n+1}}_{\mu_{n-1}\mu_{n+1}} = \sum_{\mu_n} M(n)^{i_n}_{\mu_{n-1}\mu_{n}}M(n+1)^{i_{n+1}}_{\mu_n\mu_{n+1}}.
\end{equation} 
Then one applies the two-qubit gate $U$ and obtains
\begin{equation}
(T')^{i'_{n}i'_{n+1}}_{\mu_{n-1}\mu_{n+1}} = \sum_{i_{n}i_{n+1}} U_{i'_{n}i'_{n+1},i_{n}i_{n+1}} T^{i_{n}i_{n+1}}_{\mu_{n-1}\mu_{n+1}}.
\end{equation}
In a last stage, considering the tensor $T'$ as a matrix with indices spanned by $(i'_n,\mu_{n-1})$ and
$(i'_{n+1},\mu_{n+1})$, one performs a singular value decomposition and writes
\begin{equation}
(T')^{i'_{n}i'_{n+1}}_{\mu_{n-1}\mu_{n+1}} = \sum_{\mu_n} X^{i'_n}_{\mu_{n-1}\mu_{n}} S_{\mu_{n}}Y^{i'_{n+1}}_{\mu_n\mu_{n+1}}
\label{eq:defT'}
\end{equation}
where the tensors $X$ and $Y$ are formed of orthogonal vectors while the vector $S_\mu$ contains the singular values of $T'$.
Here $S_\mu$ has up to $2\chi$ components (irrespective of the nature of the two-qubit gate) so that exact algorithms imply a doubling of $\chi$ after each application of a two-qubit gate.
In the spirit of DMRG like algorithms, we truncate $S_\mu$ and keep only its $\chi$ largest components to obtain $S'_\mu$.
The new MPS tensors are then simply given by 
\begin{eqnarray}
M'(n)^{i_n}_{\mu_{n-1}\mu_{n}} &=& X^{i_n}_{\mu_{n-1}\mu_{n}} S'_{\mu_n} \\
\label{eq:1D:Last}
M'(n+1)^{i_{n+1}}_{\mu_n\mu_{n+1}} &=& Y^{i_{n+1}}_{\mu_n\mu_{n+1}}
\end{eqnarray}
which completes the algorithm. Overall, the cost of applying a two-qubit gate is dominated by the SVD step which scales as $\chi^3$. We emphasize that such an algorithm can do anything that a quantum computer does but the reverse statement is not true: in the MPS approach, one holds the full wavefunction in memory which provides much more information than can be obtained from samples of the wavefunction. For instance, one can compute 
bipartite entanglement entropy of an MPS, and it is straightforward to calculate quantities such as observables or correlation functions without any statistical errors. The MPS format also satisfies the sample and query access criteria needed for quantum inspired de-quantizing algorithms \cite{Chia2019}.

\subsection{Random Quantum Circuit}

\begin{figure}
    \centering
    \includegraphics{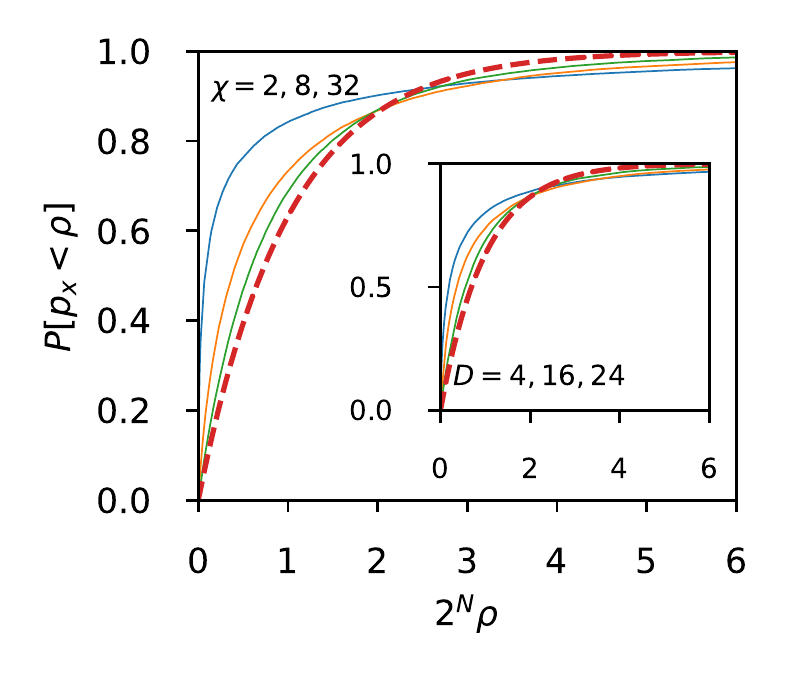}
    \caption{Cumulative distribution $P(p_x<\rho)$ where \mbox{$p_x = |\langle x|\Psi\rangle|^2$} for $N=15$. The dashed line corresponds to the Porter-Thomas distribution $P_{PT}(\rho) = 1 - (1-\rho)^{2^N-1}$. Main panel: $D=24$ and various MPS truncation levels $\chi = 2$ (blue), $8$ (orange), and $32$ (green). Inset: exact results for $D=2$ (blue), $16$ (orange), and $24$ (green)}
    \label{fig:PT}
\end{figure}

Fig.~\ref{fig:Sketch}a shows the quantum circuit used in our numerical experiments. It
consists of alternating layers of one-qubit and two-qubit gates. 
This circuit has been designed following the proposal of \cite{Boixo2018} in order to create strongly entangled states in as few operations as possible. 
It it believed to be one of the most difficult circuit to simulate on a classical computer since its many-qubit quantum state is extremely sensitive to modification of any of the gates. 
The one-qubit gates $U_n$ represented as colored squares in Fig.~\ref{fig:Sketch}a are chosen randomly such as to remove any structure or symmetry from the many qubit state. A gate $U_n$ is a rotation $U_n = \exp (-i\theta_n \vec \sigma . \vec m_n)$
of angle $\theta_n$ around a unit vector $\vec m_n = (\sin \alpha_n\cos\phi_n,\sin\alpha_n\sin\phi_n,\cos\alpha_n)$ ($\vec\sigma$ is the vector of Pauli matrices). We take the angles $\theta_n$, $\alpha_n$, and $\phi_n$ to be uniformly distributed (note that the resulting matrix $U_n$ is {\emph not} distributed according to the Haar distribution of $U(2)$).  While the $U_n$ are random, the actual sequence used is carefully 
recorded for comparison with e.g. exact calculations. 
We call the number of two-qubit gate layers applied the depth $D$ of the circuit, focusing on 
the number of two-qubit gate layers because those are the only source of imperfection in our calculations. In real quantum computers, two-qubit gates also dominate the errors over one-qubit gates in terms of fidelity. However real quantum computers also have other sources of error (decoherence,  unknown couplings between qubits, leakage to non-computational states...) not present in the algorithm. After a depth $D\sim N$, the state obtained with the circuit of Fig.~\ref{fig:Sketch}a is totally scrambled and well described by a Porter-Thomas  distribution. This is illustrated in Fig.~\ref{fig:PT} where the cumulative distribution of $p_x = |\langle x|\Psi\rangle|^2$ is compared to the Porter-Thomas form for various maximum MPS bond dimensions (main panel) and for various depths using exact calculations (inset). One indeed observes that the distribution quickly approaches the chaotic Porter-Thomas distribution as one increases the bond dimension $\chi$. 

\subsection{Effective two-qubit gate fidelity} 
Let us introduce the main quantity of interest for this study, the effective two-qubit fidelity $f_n$. The effective two-qubit fidelity $f_n$ is the computational analogue to the fidelity reported experimentally for two-qubit gates.
$f_n = 1$ for a perfect calculation, but the truncation of the MPS will induce $0<f_n<1$.

Let us call $|\Psi_T(n)\rangle$ the MPS state after a sequence of $n$ individual two-qubit gates ($n \approx (N-1) D/2$ for the circuit of Fig.~\ref{fig:Sketch}a). Up to irrelevant one-qubit gates, $|\Psi_T(n)\rangle$ is obtained by applying one Control-Z gate $C_Z$ onto \mbox{$|\Psi_T(n-1)\rangle$} followed by the truncation operation which introduces a finite error. We define the effective fidelity $f_n$ as,
\begin{equation}
f_n =  |\langle \Psi_{T}(n)|C_Z|\Psi_T(n-1)\rangle|^2 
\end{equation}
and the corresponding error rate $\epsilon_n$ as,
\begin{equation}
\epsilon_n = 1- f_n.
\end{equation}
$f_n$ can be calculated using the contraction algorithm in $N\chi^3$ operations. However, when the MPS is in canonical form, $f_n$ is simply obtained without any additional calculations as,
\begin{equation}
f_n = \left(\sum_{\mu=1}^\chi S_\mu^2 \right)/ \left( \sum_{\mu=1}^{2\chi} S_\mu^2 \right)
\label{eq:def_f_svd}
\end{equation}
where recall that $2 \chi$ is the maximum possible number of non-zero singular values of the tensor
$T'$ in Eq.~(\ref{eq:defT'}). The denominator above is always equal to one for a state which is normalized
before it is acted on by a two-qubit gate.
We have explicitly checked the equivalence between the two algorithms.

\begin{center}
\begin{figure}[h]
    \centering
    \includegraphics{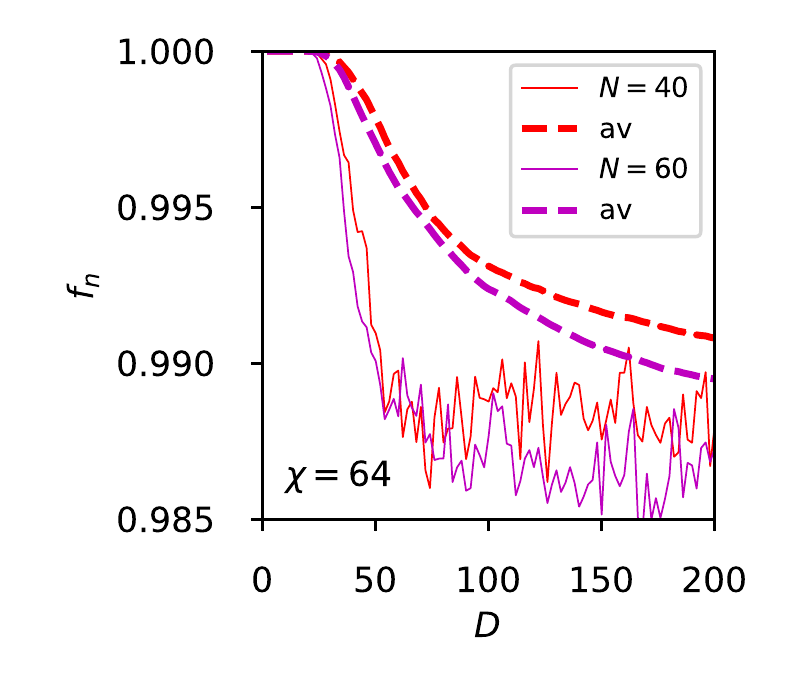}
    \caption{Effective two-qubit gate fidelity $f_n$ as a function of the depth $D$ of the circuit for $\chi = 64$ and the Control-Z gate for $N=40$ (red) and $N=60$ (magenta). The thin lines correspond to the 
 geometric average of $f_n$ over one full sequence, i.e. all the two-qubits gates performed between depth $D-2$ and depth $D$ ($N-1$ two-qubits gates). The thick dashed lines correspond to $f_{av}$, the geometric average of $f_n$ over all two-qubit gates since the beginning of the circuit up to depth $D$.}
    \label{fig:f_vs_D}
\end{figure}
\end{center}

A typical simulation is shown in Fig.~\ref{fig:f_vs_D} for the circuit with the Control-Z gate.  
At small depth $D<2\log_2\chi$, the simulation is exact and $f_n=1$. Above this threshold, one starts 
to truncate the MPS after each two-qubit gate. We observe a  transient regime where $f_n$ decreases after which 
$f_n$ quickly saturates at a constant value, here around $0.988$. 
The first thing to notice in Fig.~\ref{fig:f_vs_D} is that these simulations are many orders of magnitude easier than an equivalent {\it perfect} calculation: 
simulating the exact state for $N=60$ and $D=200$ would be out of reach even with thousand of years of computing time on the largest existing supercomputer. Yet here, these simulations of a {\it noisy} quantum computer have been performed on a laptop. The averaged fidelity for a modest $\chi=64$ is better than 99\% which already corresponds to qubits of very good quality. This is rather remarkable since the percentage of the Hilbert space spanned by the MPS ansatz is only a very tiny fraction $\sim 10^{-13}$ percent of the whole Hilbert space. After the transient regime, $f_n$ is, up to some fluctuations, independent of both $D$
and $N$. The second statement is true up to small $1/N$ corrections. These corrections arise from the fact that
the fidelity associated with gates applied on the edge of the system (i.e. associated to matrices $M(i)$ with $i<2\log_2\chi$ or $N-i < 2\log_2\chi$) is always equal to unity since the entanglement entropy associated to the subsystem of qubits $i<a$ is bounded by $S\le a \log 2$.

\begin{center}
\begin{figure}[h]
    \centering
    \includegraphics{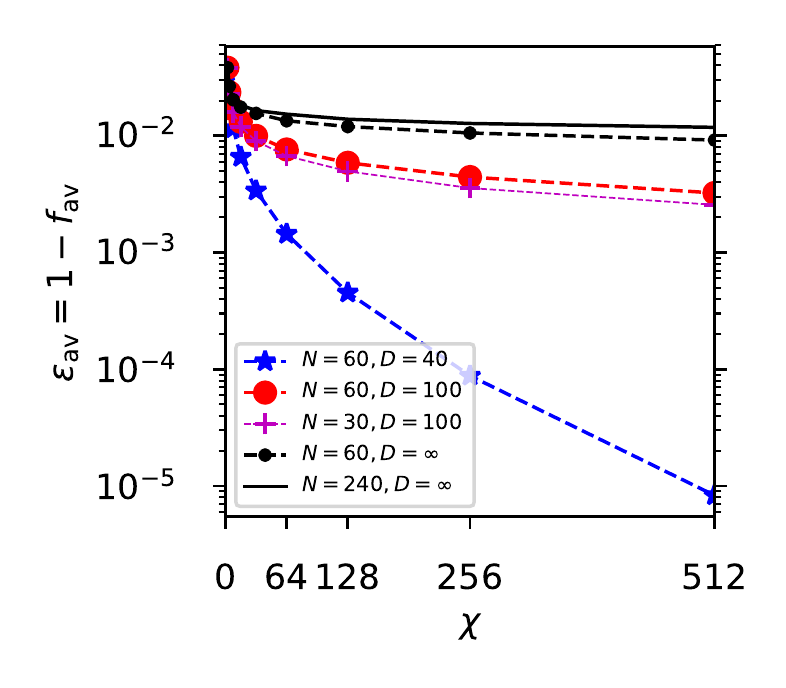}
    \caption{Geometric average of the residual error per gate $\epsilon_{\rm av} = 1- f_{\rm av}$ as a function of the bond dimension $\chi$. The average is performed over the entire circuit except for the black curves ($D=\infty$) where it is restricted to the regime where the fidelity has reached its asymptotic value ($100\le D \le 200$). For the largest system $N=240$, we have also excluded the gates on the edges of the system in our calculation as they have by construction perfect fidelity. }
    \label{fig:f_vs_chi_1D}
\end{figure}
\end{center}

Our main goal is to understand how the residual error $\epsilon_n = 1 - f_n$ decreases as one increases the bond dimension $\chi$. As $\chi$ approaches $\chi = 2^{N/2}$, one must have $\epsilon_n \rightarrow 0$. This
is because reshaping the wavefunction as a $2^{N/2} \times 2^{N/2}$ matrix implies a maximum rank of $2^{N/2}$
for any factorization of the wavefunction, thus an MPS with $\chi = 2^{N/2}$ remains exact. However, here we are interested in the regime $\chi\ll 2^{N/2}$ which remains accessible to simulations.
Fig.~\ref{fig:f_vs_chi_1D} shows how the residual error $\epsilon_n = 1 - f_n$ decreases with increasing the bond dimension. The main finding of Fig.~\ref{fig:f_vs_chi_1D} is that the residual error per gate at large depth $D$ and number of particle $N$ eventually saturates at a finite value, in this case around 
$\epsilon_\infty \approx  10^{-2}$. In other words, this algorithm can simulate any 1D quantum computer that has a two-qubit gate fidelity smaller than $f_\infty =99\%$ 
{\it at a linear cost in both $N$ and $D$}. 
As the depth or number of qubits is reduced, the average fidelity increases. The black cross in Fig.~\ref{fig:f_vs_chi_1D} corresponds to a calculation where only the last part of the circuit has been taken into account in the calculation of the average fidelity, i.e. the average is performed for $D>100$ where the system has already entered its stationary regime. Note that in that regime, there remains a small logarithmic decrease of the error: as $\chi$ increases a number $\propto \log_2 \chi$ of gates close to the edges of the system become exact, as discussed above.
The black line in Fig.~\ref{fig:f_vs_chi_1D} corresponds to calculation made in a
larger system of $N=240$ qubits where we have restricted the calculation of the fidelity to the gates for qubits in the center of the system (i.e. away from the edges where the fidelity is perfect) as well as removed the small depth regime
(only gates for $100\le D\le 200$ are taken into account). For this case, we observe a clear saturation of the error rate to a finite value $\epsilon_\infty$.
As we shall see, decreasing the error rate beyond  $\epsilon_\infty$ requires an exponential effort.

\section{Links between two-qubit and multi-qubit fidelity}
Before investigating the origin of $\epsilon_\infty$, we make a short detour to discuss how the effective
two-qubit fidelity $f_n$ is related to the actual N-qubit fidelity ${\cal F}$ of the state and is related to practical estimates of the fidelity that can be measured experimentally.
 
\subsection{Multi-qubit fidelity} 
\begin{figure}[h]
    \centering
    \includegraphics{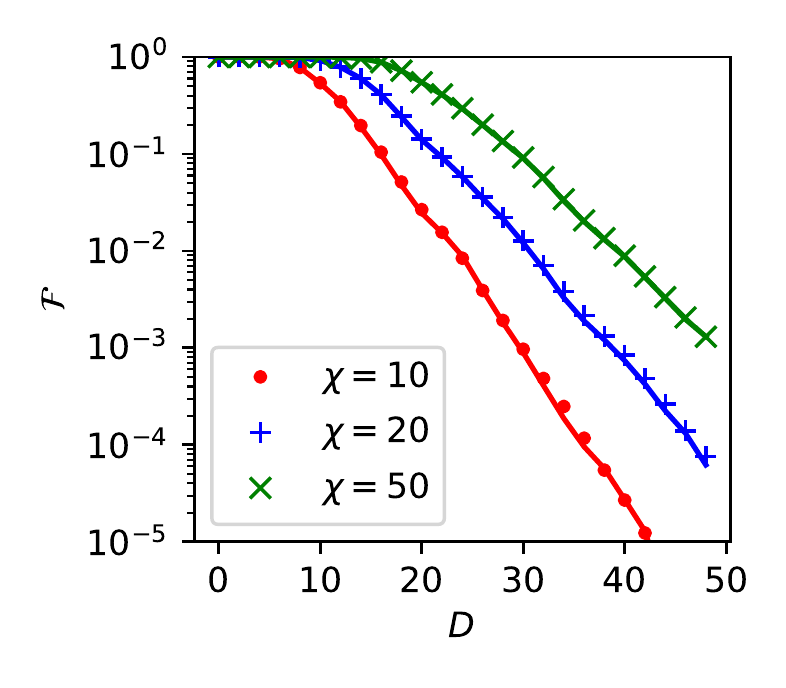}
    \caption{Fidelity ${\cal F}$ versus depth $D$ for $N=20$ and various values of $\chi= 10,20,50$. The symbols correspond to a direct calculation of ${\cal F}$ obtained by comparing with an exact calculation. The lines corresponds to the right hand side of Eq.~(\ref{estimated_fidelity}).}
    \label{fig:Fvsf}
\end{figure}

Let us call $|\Psi_P(n)\rangle$ the exact perfect state after $n$ two-qubit gates---meaning it is never truncated or otherwise approximated at any stage of its evolution by the circuit---while $|\Psi_T(n)\rangle$ is the truncated MPS state ($P$ stands for Perfect and $T$ for Truncated). The N-qubit fidelity ${\cal F}$ is defined as,
\begin{equation}
\label{eq:defF}
{\cal F}(n) =  |\langle \Psi_{P}(n)|\Psi_T(n)\rangle|^2
\end{equation}
The fidelity ${\cal F}$ is a direct measure of how reliable is our truncated state. As the errors accumulate, it is natural to expect that the fidelities $f_n$ are multiplicative,
\begin{equation}
\label{estimated_fidelity}
{\cal F}(n) \approx \prod_{i=1}^{n} f_{i}.
\end{equation}
Eq.~(\ref{estimated_fidelity}) is indeed a very accurate approximation. An analytical argument will be given below.
The validity of  Eq.~(\ref{estimated_fidelity}) can also been shown by numerical simulations.
Fig.~\ref{fig:Fvsf} shows the fidelity versus $D$ for $N=20$ particles obtained in two independent ways.
The symbols corresponds to a direct calculation of ${\cal F}$ while the lines correspond to the the right hand side of Eq.~(\ref{estimated_fidelity}). We find an almost perfect match in all the regimes that we have studied. 
Eq.~(\ref{estimated_fidelity}) is a very useful result: it relates a property of the perfect state (left hand side) to a property solely defined in terms of the MPS (right hand side). It allows us to easily estimate the fidelity in regimes where we do not have access to the exact state anymore. When $f_n$ has reached its stationary value $f_\infty$, Eq.~(\ref{estimated_fidelity}) simplifies into
\begin{equation}
\label{assymptotic_fidelity}
{\cal F}(n) \approx (f_\infty)^n \sim (f_\infty)^{\frac{ N D}{2}}.
\end{equation}
 In an actual experiment, one cannot measure the $f_n$ but rather one has access to an estimate of ${\cal F}(n)$ (see the subsection below). To compare the accuracy of the simulations with the capabilities of actual quantum chips, we therefore define the average two-qubit fidelity $f_{\rm av}$ after $n$ two-qubit gates,  
\begin{equation}
f_{\rm av} = \left(\prod_{i=1}^{n} f_{i}\right)^{1/n} \approx {\cal F}(D)^\frac{2}{ND}
\end{equation} 
where the second equality is specific to the quantum circuit studied here.

{\bf Derivation of Eq.~(\ref{estimated_fidelity})}.
Let us define a full basis of orthogonal states $|\alpha\rangle$ such that state $|1\rangle \equiv |\Psi_T(n-1)\rangle$ is our truncated state and we complement state $|1\rangle$ with an arbitrary basis. Writing $|\Psi_P(n-1)\rangle$ in that basis as $|\Psi_P(n-1)\rangle = \sum_{\alpha=1}^{2^N} p_\alpha |\alpha\rangle$, we have $p_1 = \sqrt{{\cal F}(n-1)}$. Similarly, we write $|\Psi_T(n)\rangle = \sum_{i=1}^{2^N} t_\alpha C_Z |\alpha\rangle$
with $t_1 = \sqrt{f_n}$. From these definitions, the fact that $C_Z$
is unitary and that $|\Psi_P(n)\rangle= C_Z |\Psi_P(n-1)\rangle$, we have,
\begin{equation}
{\cal F}(n) = \left[\sum_{\alpha=1}^{2^N} p_\alpha t_\alpha \right]^2= \left[\sqrt{{\cal F}(n-1) f_n} + \sum_{\alpha=2}^{2^N} p_\alpha t_\alpha \right]^2
\end{equation}
As the fidelity goes down, the $p_\alpha$ and $t_\alpha$ become increasingly decorrelated, in particular in sign. Assuming random signs between the $p_\alpha$ and the $t_\alpha$ and using that $p_\alpha \sim 1/\sqrt{2^N}$, we find that the second term in the above equation is at most of order $1/\sqrt{2^N}$ and is therefore negligible.
Eq.~(\ref{estimated_fidelity}) follows directly. 

We end this subsection by proving a weaker but exact bound for shallow circuits without the above assumption. 

The Schwartz inequality implies that,
\begin{equation}
\left(\sum_{\alpha=2}^{2^N} p_\alpha t_\alpha\right)^2 \le \sum_{\alpha=2}^{2^N} p_\alpha^2 \sum_{\alpha=2}^{2^N} t_\alpha^2 \le \epsilon_n
\end{equation}
from which we obtain,
\begin{equation}
|\sqrt{{\cal F}(n)} - \sqrt{f_{n} {\cal F}(n-1)}| \le \sqrt{\epsilon_n}
\label{eq:ex}
\end{equation}
The Eq.~(\ref{eq:ex}) bound is exact, but saturating this bound in practice implies that all the terms
$p_\alpha t_\alpha$ interfere constructively which is not realized in actual circuits. Eq.~(\ref{eq:ex}) 
implies that,
\begin{eqnarray}
{\cal F} (n) &\ge& \left( \sqrt{f_n{\cal F}(n-1)} - \sqrt{\epsilon_n}\right)^2 \nonumber \\
                    &\ge& {\cal F}(n-1) - 2\sqrt{\epsilon_n}
\end{eqnarray}
from which one can prove that, 
\begin{equation}
\label{exact_fidelity}
{\cal F}(n) \ge 1- 2\sum_{i=1}^n  \sqrt{\epsilon_i}
\end{equation}
The exact statement Eq.~(\ref{exact_fidelity}) can be useful for small depth circuits where the actual decrease of the fidelity ${\cal F}(n)$ is indeed linear with $n$, before one enters into the true exponential regime.

\subsection{Other fidelity metrics}
So far we have used the overlap ${\cal F}$ between the exact state $|\Psi_P\rangle$ and our approximate state $|\Psi_T\rangle$ as our metric for
the fidelity of the calculation. It is a natural metric as it measures the probability for the approximate state to be in the exact state one. It is bounded $0\le {\cal F} \le 1$ and is nicely related to the probabilities per gate $f_n$ through the formulas of the preceding subsection.

However ${\cal F}$ cannot be directly measured experimentally, so that other fidelity metrics must be designed. Indeed, in an actual quantum computer, the only existing output are samples of bitstrings $x=i_1i_2...i_N$ distributed  according to $|\langle x|\Psi_T\rangle|^2$. A natural metric is the cross entropy
defined as
\begin{equation}
{\cal C} = -\sum_x |\langle x|\Psi_T\rangle|^2 \log |\langle x|\Psi_P\rangle|^2
\end{equation}
Cross entropy is a standard tool of machine learning and has several interesting properties. First it is measurable through sampling as
\begin{equation}
{\cal C} = -\lim_{M\rightarrow\infty} \frac{1}{M} \sum_{m=1}^M \log |\langle x_m|\Psi_P\rangle|^2
\end{equation}
where the $x_m$ are the output of the quantum computer when the experiment is repeated $M$ times. Second, the cross entropy between two distributions $|\langle x|\Psi_T\rangle|^2$ and $|\langle x|\Psi_P\rangle|^2$ is maximum when the two distribution are identical. Hence it is a genuine measure of the likelihood of the
two distributions. Cross entropy was proposed in \cite{Boixo2018} as a fidelity metric. Note however that the cross entropy is not a symmetric function of the two distributions. In particular it is strongly affected by particular configurations $x$ where $|\langle x|\Psi_P\rangle|^2$ is very low but
$|\langle x|\Psi_T\rangle|^2$ is not. 

Cross entropy was eventually abandoned by the Google team and replaced \cite{Arute2019} by the cross entropy benchmarking (XEB) defined as
\begin{equation}
{\cal B} = -1 + 2^N \sum_x |\langle x|\Psi_T\rangle|^2 |\langle x|\Psi_P\rangle|^2
\end{equation}
XEB is also sampleable and is symmetric with respect to the two distributions.
When the approximate state is the uniform distribution, the XEB metric vanishes, ${\cal B}=0$ indicating a total lack of fidelity. However, when the approximate state is actually exact, the value of the XEB metric can be arbitrary. When the approximate state is exact {\it and} distributed according to the Porter Thomas distribution (which happens in our circuits after a few cycles), then the XEB metric gets a well defined ${\cal B}=1$ value. The XEB metric is not in general a good measure of the likelihood between two distributions: for a given perfect state, it is maximum when the approximate state is sharply peaked around the values of $x$ where the perfect state is maximum. In our circuit the initial
value of XEB is exponentially high ${\cal B}=2^N -1$ and quickly decreases as the distribution approaches the Porter-Thomas one. Calling $D^*$ the depth after which XEB has reached unity (ideally $D^*$ would the depth after which $|\langle x|\Psi_P\rangle|^2$ corresponds to Porter-Thomas), we find empirically that 
\begin{equation}
\label{eq:xeb_vs_F}
{\cal F}_n \approx {\cal F}(D^*) {\cal B}_n
\end{equation}
Equation (\ref{eq:xeb_vs_F}) could be used to estimate the actual fidelity
${\cal F}$ from XEB measurements.

\begin{figure}
    \centering
    \includegraphics{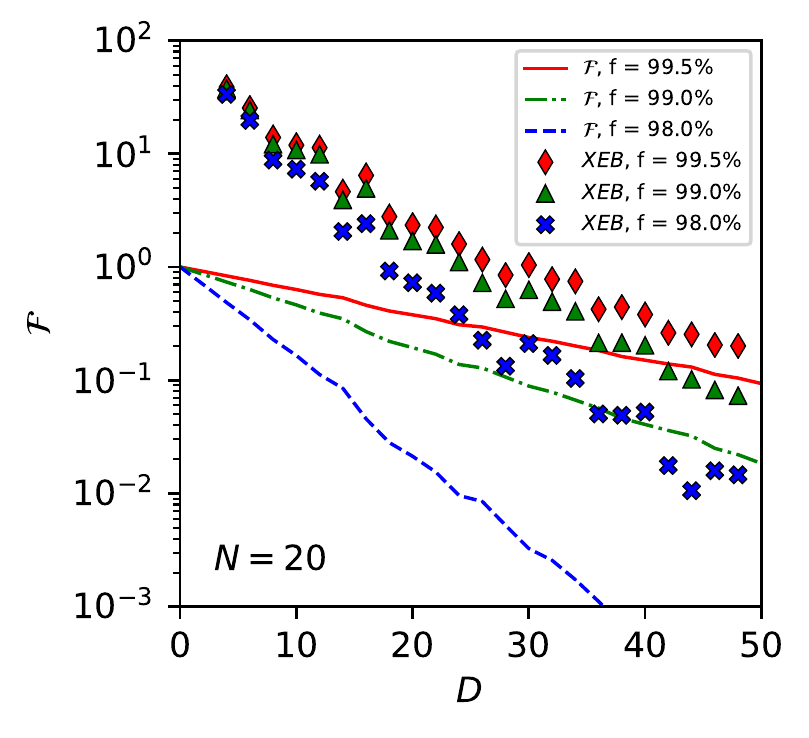}
    \caption{Comparison between the fidelity ${\cal F}$ (lines) and the XEB metric ${\cal B}$ (markers) as a function of depth $D$. Different colors label different levels of noise on the two-qubit gates, respectively $f=99.5\%$ (red), 
    $f=99\%$ (green) and $f=98\%$ (blue). The calculations were performed for the 1D random circuit with $N = 20$ qubits. 
    }
    \label{fig:xeb_vs_f}
\end{figure}

Figure \ref{fig:xeb_vs_f} show an example of calculations contrasting the fidelity ${\cal F}$ with the XEB metric. Here we have used no truncation but added some noise on the two qubit gate so as to induce a finite fidelity per gate $f$. We find that both ${\cal F}$ and XEB decay exponentially with consistent decay rates. However, the large difference of the initial values 
at $D=0$ leads to a shift of the fidelity which is significantly lower than the XEB curve. This shift increases as the fidelity is lowered and corresponds typically to one order of magnitude for a typical experimental value $f=99\%$.

\section{Random Tensor Theory of $\epsilon_\infty$}
We now turn back to the discussion of the asymptotic value $f_\infty$ reached by the two-qubit gate fidelity in our calculations. The first remark of importance is that $f_\infty$ is a property associated with a single tensor of the full MPS state: if we apply a gate between qubit $i$ and qubit $i+1$, only the associated $T'$ tensor defined in Eq.~(\ref{eq:defT'}) comes into play. Since the whole goal of our quantum circuit is to scramble the wavefunction as efficiently as possible, a natural hypothesis is that the tensors $M(i)$ and $M(i+1)$ become eventually 
well described by totally random tensors. In this section we explore this possibility and calculate the properties of the associated tensor $T'$ as well as the corresponding two-qubit gate fidelity $f_{\rm GTE}$. We find that 
the distribution of singular values of $T'$ obtained from the random ensemble closely matches what we observe in the MPS state.

In the spirit of random matrix theory~\cite{Mehta2004,Beenakker1997}, we introduce the Gaussian tensor ensemble (GTE) where a tensor $M_{\mu\nu}^{i}$ is supposed to be totally random. The GTE can be thought of as a ``worse case scenario'' where the quantum circuit is so chaotic that the tensors are left with no structure.
In the GTE, the tensor $M$ are distributed
according to 
\begin{equation}
P\left[M_{\mu\nu}^{i}\right] \propto \exp\left[ -\frac{1}{2} \sum_{\mu\nu i} |M_{\mu\nu}^{i}|^2\right]
\end{equation} 
where the sum over $\nu$ spans $1\dots\chi$, the sum over $i$ spans $0,1$ and the sum over $\mu$ span $1\dots\beta\chi$. In the remaining of this section, we restrict ourselves to $\beta=1$ which corresponds to the tensors of Eq.(\ref{eq:mps}). We shall have an example of $\beta=2$ for the grouped-qubit algorithm we will discuss in section \ref{algorithms beyond eps infty}.
From two such tensors, we apply a two-qubit gate following Eq.(\ref{eq:1D:1})-(\ref{eq:1D:Last})
constructing the associated tensor $T$ and $T'$ and the SVD of $T'$.
From the $2\beta\chi$ singular values $S_\mu$ of $T'$, we can obtain the associated fidelity $f_{\rm GTE}$ through Eq.~(\ref{eq:def_f_svd}).
\begin{figure}
    \centering
    \includegraphics{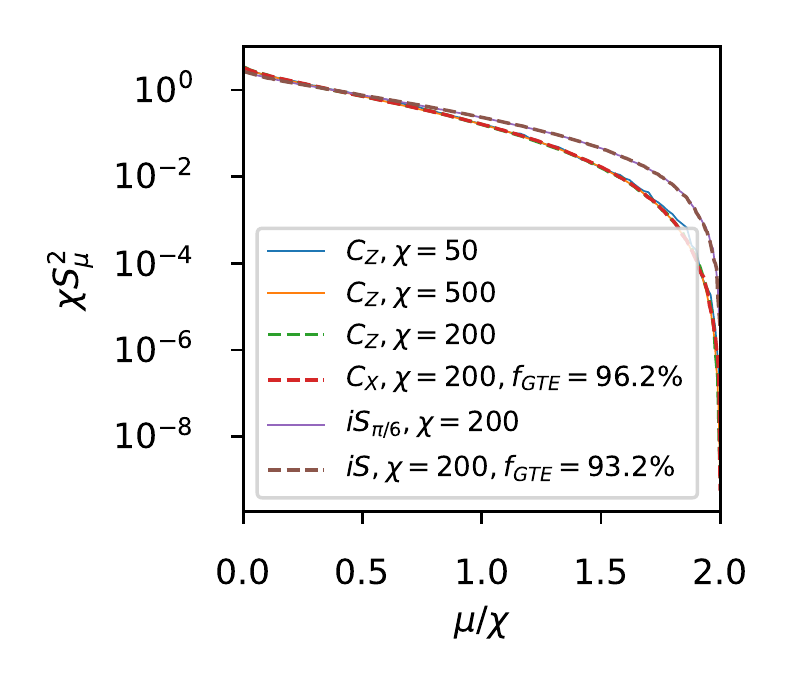}
    \caption{Squared singular values $S_\mu^2$ of the matrix $T'$ obtained from the GTE ensemble. We find a perfect scaling of the form $S_\mu^2 = g(\mu/\chi)/\chi$ where $\mu$ is the index of the $\mu^{\rm th}$ singular value. The two bundles of curves correspond respectively to the $C_X$,$C_Z$ gates (two non-zero eigenvalues) and the $iS_{\pi/6}$/$iS$ gates (four non-zero eigenvalues). Within one bundle, the different curves are indistinguishable.}
    \label{fig:gte}
\end{figure}

Fig.~\ref{fig:gte} studies the distribution of the singular values $S_\mu$ for tensor $T'$ obtained from the GTE.
The singular values are sorted in order of decreasing magnitude and plotted as a function of the 
index $\mu=1,\ldots,2\chi$. Plotting $\chi S_\mu^2$ as a function of $\mu/\chi$, we observe that all the different values of $\chi$ collapse onto a single curve. In other words, we find that there is some function $g(x)$ such that
\begin{equation}
S_\mu^2 = \frac{1}{\chi}g\left(\frac{\mu}{\chi}\right).
\end{equation}
This scaling is already valid for rather small values of $\chi$. This observation can probably be put on firm mathematical grounds - it is consistent with the usual scaling of the semi-circular law of the so-called Gaussian unitary ensemble - but for the moment it is merely an empirical statement made from numerical evidence. It follows from this scaling that $f_{\rm GTE}$ very quickly converges to
\begin{equation}
f_{\rm GTE} = \frac{\int_0^1 dx\ g(x)}{\int_0^{2\beta} dx\ g(x)}.
\end{equation}
In other word, one finds a finite value of the fidelity that is independent of $\chi$. The resulting $f_{\rm GTE}$ depends on the other hand on the two-qubit gate used. Control-Z ($C_Z$) and control-NOT ($C_X$) are equivalent (they are related to each other through a change of basis of the second qubit) and corresponds to $f_{\rm GTE}=96.2\%$. Gates like the iSWAP gate ($iS$) or iSWAP followed by a $\pi/6$ rotation over the z-axis ($iS_{\pi/6}$, close to what is used in \cite{Arute2019}) have 4 different singular values which roughly doubles the error with respect to $C_Z$
($f_{\rm GTE}=93.2\%$).

\begin{figure}
    \centering
    \includegraphics{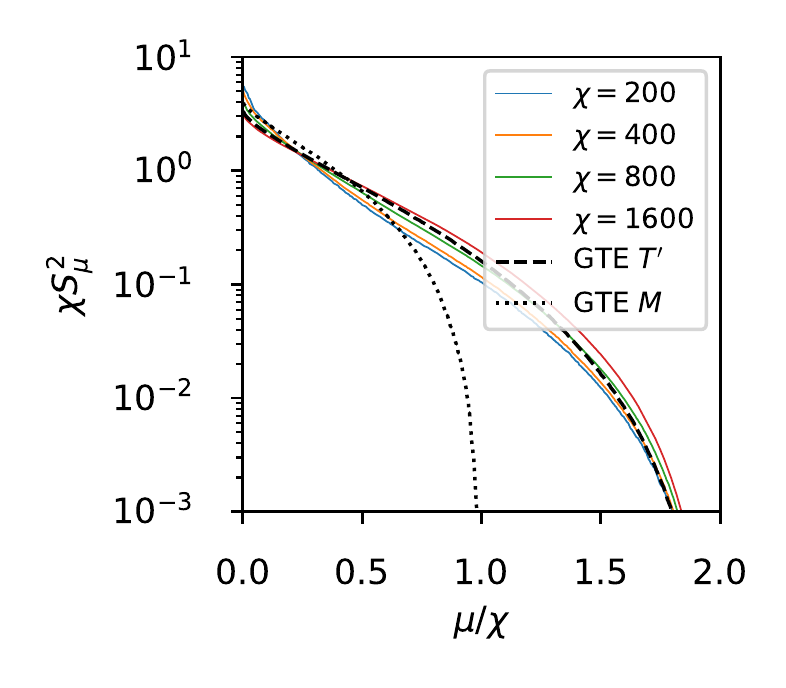}
    \caption{Squared singular values $S_\mu^2$ of the matrix $T'$ obtained from the MPS simulations of $N=30$ qubits and a depth of $D=60$ for various values of $\chi$. The singular values correspond to a gate $C_X$ performed in the middle of the system. Dotted line: Squared singular values of the $M$ matrix in in the GTE. Dashed line: Squared singular values of $T'$ in the GTE.
    }
    \label{fig:gte_vs_mps}
\end{figure}

Fig.~\ref{fig:gte_vs_mps} shows how the distribution of the singular values in the GTE compares to the one obtained in 
the MPS simulation. We find a close agreement between GTE and the MPS simulations when looking at the $T'$ tensor
for a gate in the center of the system and at large depth. The agreement is not perfect however, and we observe that the asymptotic fidelity of MPS simulations is always better than the one found in GTE, 
\begin{equation}
\label{eq:boundGTE}
f_\infty \ge  f_{\rm GTE}.
\end{equation}
To try and understand why the inequality in Eq.~(\ref{eq:boundGTE}) is not saturated, we plot
in Fig.~\ref{fig:gte_vs_mps} the distribution of the singular value of the initial tensor $M$
(dotted line). After truncation, the distribution of the singular values of $M'$ are given by
the dashed line restricted to $0\le \mu/\chi \le 1$ (up to a small shift due to the normalization of the state). These two distributions differ very significantly. In order to saturate the bound
of Eq.~(\ref{eq:boundGTE}) we would need extra steps to scramble the distribution of $M'$ back to
the distribution of $M$ (i.e. go from the dashed line to the dotted line). However, since in our protocol only a single one-qubit gate separates one truncation from  the next one, we find that it is not sufficiently chaotic and therefore we never reach the ``worse case scenario'' of the GTE.

To summarize, $f_{\rm GTE}$ can be thought as a lower bound for the fidelity found in the simulations for large enough $\chi$ (typically $\chi\ge 300$ in practice) and large enough depth. Getting beyond the asymptotic value requires algorithms that have an exponential cost. In the following section we describe possible strategies. 

\section{Algorithms for getting beyond $\epsilon_\infty$} \label{algorithms beyond eps infty}
The algorithm discussed above can also be used for 2D arrays, since any two qubit gates
between distant qubits can always be written as a combination of gates on neighboring qubits using SWAP gates. However, this is inefficient and leads to a decrease of the effective $f$ as the transverse dimension of the 2D array increases. Another limitation of the above algorithm is that one cannot efficiently simulate systems that have a fidelity above $f_\infty$. 

There are multiple strategies that could be used to go beyond the above algorithm. In particular, recent progress in the algorithms for contracting tensor networks, such as \cite{Pan2019} could be
interesting candidates in 2D. Below, we follow  a very simple strategy where we keep using MPS states,
but group the qubits so that each tensor now represents several qubits. 

\subsection{Grouped MPS State and Extraction Algorithm \label{sec:grouped}}
We now consider the MPS structure sketched in Fig.~\ref{fig:Sketch}c where each tensor addresses several qubits. 
We now have $P \le N$ tensors $M(n)$ each addressing $N_n$ qubits with $\sum_{n=1}^P N_n = N$.
The tensors $M(1)$ and $M(P)$ possess $N_n+1$ indices while the others possess $N_n+2$ indices,
\begin{equation}
M(n)_{\mu\nu}^{i_1i_2...i_{N_n}}
\end{equation}
The number of elements of these tensors is $\chi^2 2^{N_n}$ so that the computing time now increases exponentially with the number of qubits per tensor. On the other hand, the two qubit gates that are performed inside a given tensor $M(n)$ are now handled exactly, so that the average fidelity of a circuit increases.

\begin{figure}
    \centering
    \includegraphics[width=0.95\columnwidth]{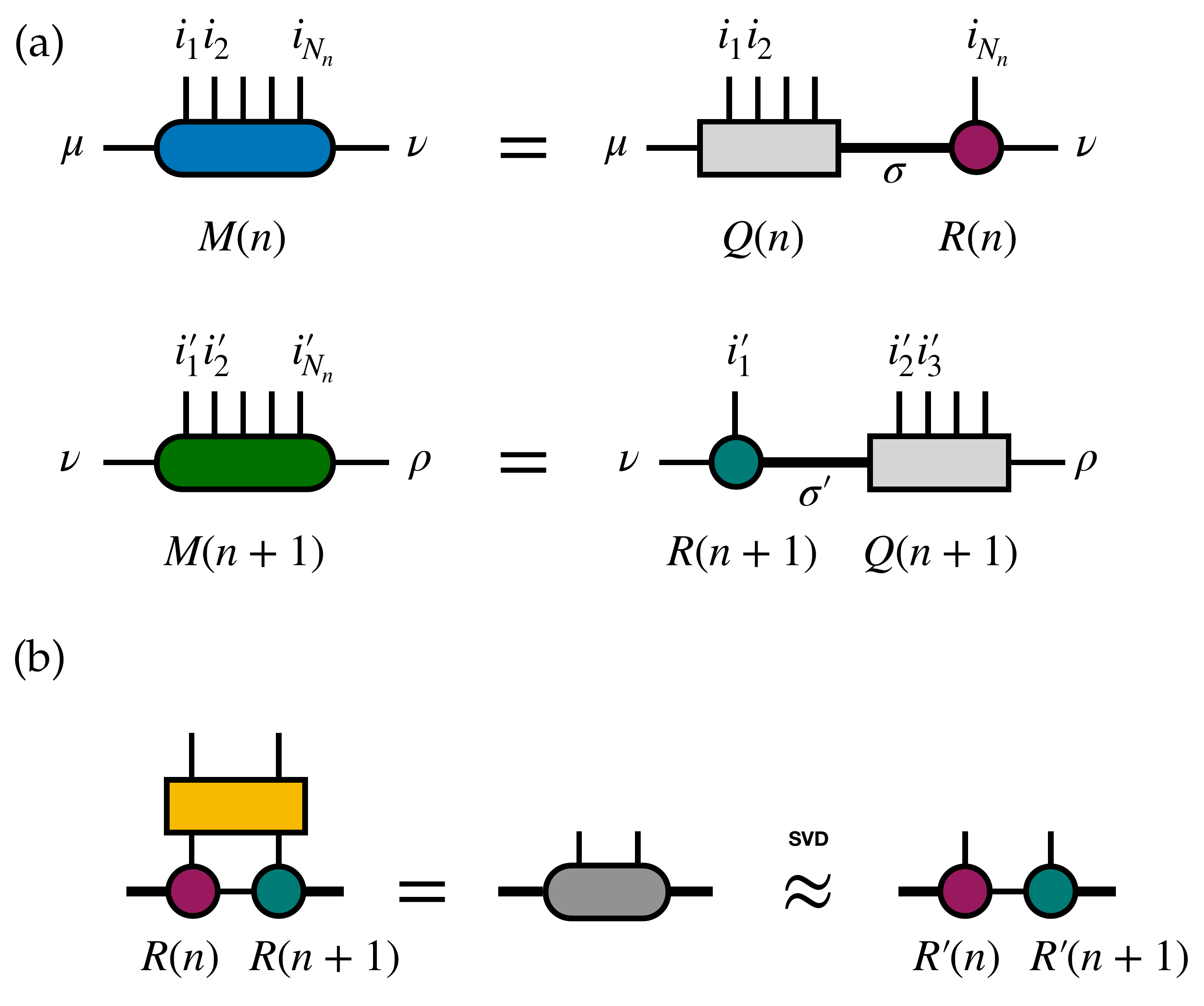}
    \caption{Main steps for applying a gate which acts across two grouped MPS tensors, as described in Eqs.~(\ref{eqn:mq1})--(\ref{eqn:mq2}). In (a) the grouped MPS tensors $M(n)$ and $M(n+1)$ are 
    exactly factorized using QR decompositions, such that the $R(n)$ and $R(n+1)$ tensors carry the qubit
    indices acted on by the gate and the newly introduced indices $\sigma$ and $\sigma'$ range over $2\chi$
    values. In (b) the gate acts on the product of $R(n)$ and $R(n+1)$, and the resulting
    tensor is factorized using an SVD truncated to $\chi$ singular values. Finally, to update the
    MPS (not shown), one computes the new tensors $M'(n)=Q(n)R'(n)$ and $M'(n+1)=R'(n+1)Q(n+1)$ which
    diagrammatically looks like step (a) but in reverse.
    } 
    \label{fig:gmps_diagrams}
\end{figure}

To perform a two-qubit gate between neighboring tensors $M(n)$ and $M(n+1)$, one proceeds in three steps.
The first two are shown diagrammatically in Fig.~\ref{fig:gmps_diagrams}.
In the first step, one performs a $QR$ decomposition of the two tensors to ``extract'' smaller tensors corresponding to the involved qubits. Assuming (without loss of generality) that the two qubit gate involves qubit $N_n$ of tensor $M(n)$ and qubit $1$ of tensor $M(n+1)$, one decomposes $M(n)$ as
\begin{equation} \label{eqn:mq1}
M(n)_{\mu\nu}^{i_1i_2...i_{N_n}} =  \sum_{\sigma=1}^{2\chi} Q(n)_{\mu,\sigma}^{i_1 i_2...i_{N_n-1}} R(n)_{\sigma,\nu}^{i_{N_n}} 
\end{equation}
where the ``vectors'' of $Q(n)$ indexed by $\sigma$ are orthonormal. The important point here is that the index $\sigma$ takes only $2\chi$ values.
Similarly, we write:
\begin{equation}
M(n+1)_{\nu\rho}^{i'_1 i'_2...i'_{N_{n+1}}} =  \sum_{\sigma=1}^{2\chi}  R(n+1)_{\nu,\sigma'}^{i'_1}  
Q(n+1)_{\sigma',\rho}^{i'_2...i'_{N_{n+1}}} \label{eqn:mq2}
\end{equation}
The second step follows Eqs.~(\ref{eq:1D:1})-(\ref{eq:1D:Last}) of the algorithm of Section \ref{sec:1D}
with the replacement $M(n)\rightarrow R(n)$ and $M(n+1)\rightarrow R(n+1)$, and is shown 
for the present case in Fig.~\ref{fig:gmps_diagrams}(b). In the last step the new
tensors $M'(n)$ and $M'(n+1)$ are obtained by contracting $Q(n)$ with $R'(n)$ and  $R'(n+1)$ with $Q(n+1)$.

The main difference between the algorithm of Section \ref{sec:1D} and the grouped MPS algorithm is that
the resulting tensor $T'$ of Eq.~(\ref{eq:defT'}) now has $4\chi$ singular values instead of $2\chi$.
As a result, upon truncation to keep only $\chi$ singular values, we anticipate that the fidelity
per gate will be smaller than in the 1D case. However, as we shall see, this decrease will be more than compensated by the gain of having perfect gates within one tensor. In the terminology of
random tensors, the grouped MPS algorithm corresponds to $\beta=2$. For the $C_Z$ gate, the GTE fidelity
drops from $f_{\rm GTE}(\beta=1)=96.2\%$ down to $f_{\rm GTE}(\beta=2)=87.4\%$.

\subsection{Application to a two dimensional circuit}
\begin{figure}
    \centering
    \includegraphics[width=0.7\columnwidth]{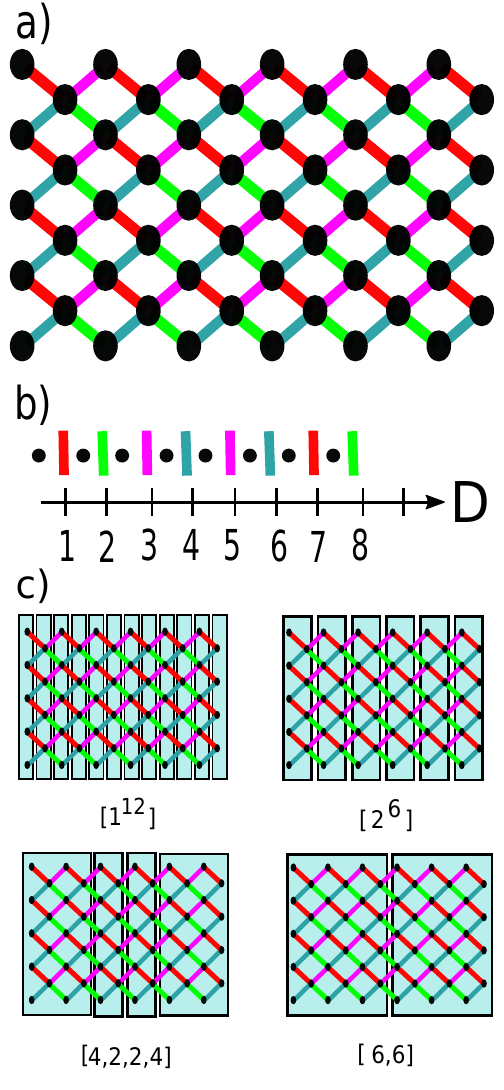}
    \caption{a) Sketch of the quantum circuit with 54 qubits in a 2D grid. The qubits are represented by the black dots while the two-qubit gates by the color links. b) The circuit alternates one-qubit gates (black dots) with two-qubit gates (here the Control-Z gate). The depth $D$ counts the number of two-qubit gates per qubit. c) Different grouping strategies for the group MPS algorithm. $[1^{12}]$ corresponds to a grouping in $12$ blocks counting $1$ column each; $[4,2,2,4]$ corresponds to a grouping in 4 blocks counting respectively $4$, $2$, $2$, and $4$ columns.}
    \label{fig:Sketch2D}
\end{figure}

We now show the results of simulations performed on a 2D circuit. To put the results into the perspective of what can be achieved experimentally, we choose a circuit very close to the one
used by the Google team in their ``supremacy'' experiment \cite{Arute2019}. We consider a 2D grid
of $54$ qubits as shown in Fig.~\ref{fig:Sketch2D}a. The circuit is shown in Fig.~\ref{fig:Sketch2D}b
and alternates one-qubit gates applied to each qubit (same distribution as in the 1D case) with two-qubits gates (Control-Z) applied on different pairs of qubits according to the color shown. Except for the choices of one- and two-qubit gates, and the number of qubits ($53$ versus $54$), the setup is identical to the ``supremacy sequence'' of the Google experiment \cite{Arute2019}. In Ref.\cite{Arute2019} a XEB fidelity ${\cal B}=0.002$ was reached after a depth $D=20$ corresponding to a 
total of $430$ two-qubit gates. Ignoring the difference between XEB and the fidelity ${\cal F}$,
this translates into $\epsilon_{\rm av}=1.4\%$ which we shall use as our reference value to evaluate 
the performance of the grouped MPS algorithm.

\begin{center}
\begin{figure}
    \centering
    \includegraphics[width=0.45\textwidth]{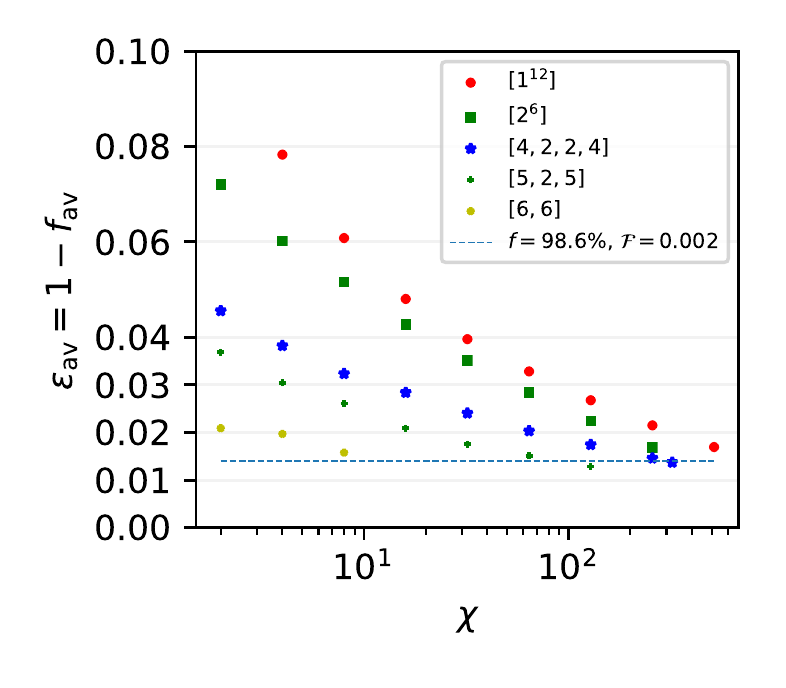}
    \caption{Residual error per gate $\epsilon_{\rm av} = 1- f_{\rm av}$ as a function of the bond dimension $\chi$ for the $2D$ circuit of Fig.~\ref{fig:Sketch2D} for a depth $D=20$. 
    The different curves correspond to different groupings. 
    The horizontal dashed line corresponds to the error rate associated with a global fidelity ${\cal F}=0.002$.}
    \label{fig:f_vs_chi_2D}
\end{figure}
\end{center}

Fig.~\ref{fig:Sketch2D}c shows various strategies for grouping the qubits. The $[1^{12}]$ grouping corresponds to $12$ tensors that contains one column of qubit each (i.e. alternatively $5$ and $4$ qubits). The $[6,6]$ grouping is the most expensive computationally with two tensors of $27$ qubit each.
Note that the tensors on the edges are less computationally costly than the middle ones, since they only have one bond index. The results of the simulations are shown in Fig.~\ref{fig:f_vs_chi_2D} for a depth of $D=20$. While the error rate is significantly larger than in the 1D case, we find that it can be brought down to less than $1.4\%$ (which corresponds to a global fidelity of ${\cal F}=0.002$) on a single core computer. The computing times of the data points of Fig.~\ref{fig:f_vs_chi_2D} range from a few seconds to less than 48 hours for the most expensive points on a non-parallel code (single core calculation). We find that the grouping strategy is effective, but not as efficient as the maximum gain that one could expect: even though some of the gates become perfect upon grouping, we observe a decrease of the fidelity
for the noisy gates which reduces the overall gain.
For $\chi=320$ and the $[4,2,2,4]$ partition where the final fidelity is slightly better than ${\cal F}=0.002$ (see Fig.~\ref{fig:f_vs_chi_2D}), the memory footprint of the calculation is 4.5 GB of memory
which represents only $1.5\times10^{-6}$ percent of the size of the total Hilbert space spanned by the $2^{54}$ qubits. 

\subsection{Split-and-Merge algorithm for more complex gates}
We end this article with results in a configuration that closely matches
the one of Ref.~\onlinecite{Arute2019}. The one-qubit gates are chosen at random between
$\sqrt{X}$, $\sqrt{Y}$ and $\sqrt{W}$ while the two-qubit gate $iS_\theta$ is a combination of iSWAP followed by a controlled rotation  along the $z$ axis
\begin{equation}
iS_\theta =
\left(
\begin{array}{cccc}
 1 & 0 & 0 & 0 \\
 0 & 0 & -i & 0 \\
 0 & -i & 0 & 0 \\
 0 & 0 & 0 & e^{-i\theta} 
\end{array}
\right)
\end{equation}
This gate has four different singular values and is therefore expected to produce more entanglement than the $C_Z$ gate. The link between number of singular values and the actual growth of entanglement is not totally straightforward, however. Indeed, the pure iSWAP gate has four different singular values $\pm 1$ and $\pm i$; yet as it preserves the structure of product states, it is trivial to simulate with perfect fidelity. In what follows, we use $\theta=1$ which is non-trivial to simulate.

The algorithm of the previous subsection behaves rather poorly for the $iS_\theta$ gate. For instance, for 
\mbox{$\chi= 128$}, and the $[4,2,2,4]$ grouping, the two-qubit gate fidelity drops from \mbox{$f\approx 98\%$} ($C_Z$) to \mbox{$f\approx 92\%$} ($iS_\theta$). However, a simple modification of the algorithm allows one to recover a much higher fidelity $f\approx 95\%$.

\begin{figure}
    \centering
    \includegraphics[width=0.8\columnwidth]{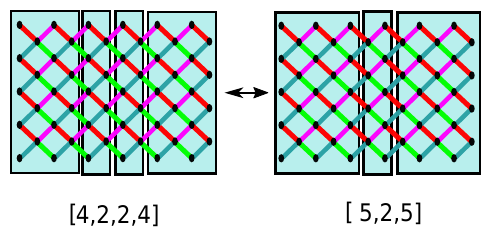}
    \caption{Schematic of the Split-and-Merge algorithm for the $[4,2,2,4]\leftrightarrow [5,2,5]$. The two-qubit gates shown in red and dark green are performed in the $[4,2,2,4]$ configuration and one switches to the $[5,2,5]$ to perform the light green and purple gates.}
    \label{fig:SaM}
\end{figure}

To study $iS_\theta$, we therefore switch to a ``Split-and-Merge" strategy: instead of ``extracting" qubits one-by-one to perform two-qubit gates as in Section \ref{sec:grouped}, we extract one full column of qubits at a time. In the Split-and-Merge strategy, we use two different groupings of the qubits, for instance 
switching between the $[4,2,2,4]$ grouping and the $[5,2,5]$ grouping  (hereafter referred to as
the $[4,2,2,4]\leftrightarrow [5,2,5]$ grouping strategy).
Switching from one grouping to another induces truncation errors. However, once the switching has been done, many two-qubit gates can be performed exactly. A schematic of the Split and Merge strategy is shown in 
Fig.~\ref{fig:SaM}
for the $[4,2,2,4]\leftrightarrow [5,2,5]$ case.

\begin{center}
\begin{figure}
    \centering
    \includegraphics[width=0.45\textwidth]{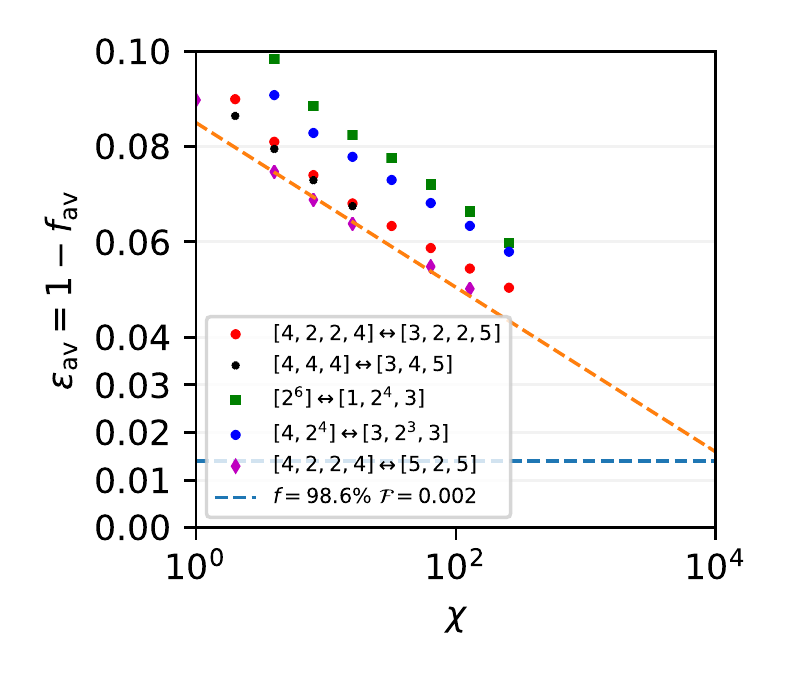}
    \caption{Residual error per gate $\epsilon_{\rm av} = 1- f_{\rm av}$ as a function of the bond dimension $\chi$ for the $iS_\theta$ gate for a 2D circuit with $N=54$ qubits and a depth $D=20$. 
    The different curves correspond to different groupings. 
    The horizontal dashed line corresponds to the error rate associated with a global fidelity ${\cal F}=0.002$.
    The orange line is just a guide to the eye.}
    \label{fig:f_vs_chi_XYW}
\end{figure}
\end{center}

Fig.~\ref{fig:f_vs_chi_XYW} shows our numerical results for $\epsilon_{\rm av}$ versus $\chi$. The curves are very similar to those obtained for $C_Z$ at similar computational cost, but with an error rate roughly three times larger than with $C_Z$.

To conclude this section, we have shown that for the Control-Z gate a simple grouping strategy allows one to reach the same fidelity as the Google experiment \cite{Arute2019} in a matter of hours on a single core computer (i.e. $f_{\rm av}\ge 98.6\%$). For the more challenging $iS_\theta$ gate, this fidelity drops down to $95\%$ for similar computing time. 

A natural question that arises is whether these algorithms may be used to defeat the claim of quantum supremacy put forward in \citep{Arute2019}, i.e. raise the fidelity from $95\%$ to $>98\%$. We have not be able to do so on a single core implementation.
However, the Split-and-Merge is to a large extent trivially parallelizable since most tensor operations contain ``spectator" indices whose different values can be fixed, and the resulting tensor ``slices'' dispatched to different computing cores or nodes. Extrapolations from our results suggest that such a parallel implementation should be able to reach fidelities in the $98-99\%$ range with a few hundred cores and a few terabytes of memory. However, such a calculation has not be attempted at the moment. 
Let us note, in any case, that not too much emphasis should be put on quantum supremacy by itself. It is not because a task is difficult to simulate that it provides a useful output. Also, there is no question that quantum many-body problems are extremely difficult to simulate. The insight that we get from the present work is an estimate of the relation between the accuracy reached in the quantum state and the underlying amount of entanglement that could potentially be exploited.

\section{Discussion}
In this work, we have discussed a practical algorithm that allows to simulate a quantum computer in a time which grows linearly with the number of qubits $N$ and the depth $D$ at the cost of
having a finite fidelity $f$ per two-qubit operation. The fidelity $f$ can be increased at a polynomial cost up to a finite value $f_\infty$; increasing it further has an exponential cost in the fidelity. Our main observation is that fidelities of the order of $99\%$, which are typical fidelities found in state of the art experiments, can be reproduced at a moderate computational cost. 

Is a fidelity of $99\%$ large or small? From an experimental physics perspective, it is certainly quite an achievement to keep several dozen qubits at this level of fidelity. From a quantum information and classical algorithms point of view, a question is what is the level of entanglement---hence the actual fraction of the Hilbert space that can truly been accessed---associated with this level of fidelity. Our MPS ansatz can provide an estimate (or at least an upper bound for one may come up with better algorithms) for this fraction. Since the MPS ansatz only spans a very tiny fraction of the overall Hilbert space, it follows that the computational power associated with fidelities in the $99\%$ range is much more limited than the full size $2^N$ of the Hilbert space would suggest. We conclude that increasing the computational power of a quantum computer will primarily require increasing the fidelity/precision with which the different operations are performed \cite{Waintal2019}.  Secondarily, one could try to improve its connectivity with e.g. quantum buses \cite{Bauerle2018} as we have seen that 1D simulations are far easier than 2D ones. However, increasing the number of qubits will remain ineffective until better fidelities have been reached.

As a side comment, our approach could also be used to get lower bounds for quantum error correction (QEC) schemes \cite{Nielsen2000}. Suppose that for a certain connectivity, one has an algorithm that can reach a fidelity $f$ in polynomial time in $N$ and $D$. Then, it is reasonable to expect that any QEC code has a threshold $p>f$. If it were not the case, one could build
a logical quantum computer with a classical one at a polynomial cost by simply simulating the QEC protocols on the classical computer. In this respect, extending our approach to a truly 2D algorithm (beyond the quasi-1D one discussed in this article) would be particularly interesting. Indeed, 2D surface codes have a particularly low threshold $p \approx 99\%$. How close to \mbox{$f=99\%$} can one get at a polynomial cost in 2D is currently an open question.

Finally, it would be interesting to perform a similar study, but of how well MPS of practical
sizes can approximate circuits designed for \emph{useful} tasks. Goals could include estimating minimum fidelities needed to perform these tasks with a high success probability and understanding crossovers where useful quantum algorithms begin to offer advantages over classical approaches.

\section*{Acknowledgments} 

XW and YZ thank the Flatiron CCQ where this work was initiated during summer 2019. XW acknowledges funding from the French ANR QCONTROL and the E.U. FET open UltraFastNano. Numerical results involving MPS were obtained using the ITensor library \cite{ITensorJulia}. The Flatiron Institute is a division of the Simons Foundation. Thanks to Thomas Ayral for interesting discussions.

\bibliography{bibliography}

\end{document}